\documentclass[12pt]{article}%
\usepackage{amsmath,latexsym}
\usepackage{graphicx}
\usepackage{amsmath}
\usepackage{amsfonts}
\usepackage{amssymb}%
\begin{document}

\title{{Exactly solvable wormhole and cosmological models
    with a barotropic equation of state }}
   \author{
  Peter K.F. Kuhfittig\\  \footnote{kuhfitti@msoe.edu}
 \small Department of Mathematics, Milwaukee School of
Engineering,\\
\small Milwaukee, Wisconsin 53202-3109, USA}

\date{}
 \maketitle

\begin{abstract}\noindent
An exact solution of the Einstein field
equations given the barotropic equation of
state $p=\omega\rho$ yields two possible
models: (1) if $\omega<-1$, we obtain the
most general possible anisotropic model
for wormholes supported by phantom
energy and (2) if $\omega>0$, we obtain
a model for galactic rotation curves.
Here the equation of state represents
a perfect fluid which may include dark
matter.  These results illustrate the
power and usefulness of exact solutions.
\\

\noindent
PAC numbers: 04.20.Jb, 04.20.-q, 04.20.Gz
\end{abstract}

\section{Introduction}\label{E:introduction}

A challenging problem in the general theory of
relativity is finding exact solutions of the
Einstein field equations.  While a solution
does not have to be exact to be valid, finding
an exact solution does have one distinct
advantage: it often yields physical insights
or unexpected connections that a numerical
solution cannot.  This paper offers an extreme
example of an exact solution that models two
completely diverse structures, traversable
wormholes and dark-matter models for galactic
rotation curves.  The latter case even suggests
that a constant tangential velocity could have
been anticipated based on the Einstein field
equations.

For the former case let us recall that
wormholes are handles or tunnels in spacetime
connecting different regions of our Universe or
different universes altogether.  That wormholes
could be actual physical structures suitable for
interstellar travel was first proposed by Morris
and Thorne \cite{MT88}.  For the wormhole
spacetime they assumed the following static
spherically symmetric line element
\begin{equation}\label{E:line1}
ds^{2}=-e^{2\Phi(r)}dt^{2}+\frac{dr^2}{1-b(r)/r}
+r^{2}(d\theta^{2}+\text{sin}^{2}\theta\,
d\phi^{2}),
\end{equation}
using units in which $c=G=1$.  Here $\Phi=
\Phi(r)$ is called the \emph{redshift function},
which must be everywhere finite to avoid an
event horizon.  The function $b=b(r)$ helps
determine the spatial shape of the wormhole
and is therefore called the \emph{shape
function}.  The spherical surface $r=r_0$ is
the \emph{throat} of the wormhole and must
satisfy the following conditions: $b(r_0)=r_0$,
$b(r)<r$ for $r>r_0$, and $b'(r_0)<1$, now
usually called the \emph{flare-out condition}.
This condition refers to the flaring out of
the embedding diagram pictured in Ref.
\cite{MT88}.  The flare-out condition can
only be satisfied by violating the null
energy condition.

An apparently unrelated topic is the existence
of galactic rotation curves.  Here we need to
recall the well-known problem that rotation
curves of neutral hydrogen clouds in the outer
regions of the galactic halo cannot be explained
in terms ordinary luminous matter.  This
phenomenon has led to the hypothesis that
galaxies and even clusters of galaxies are
pervaded by \emph{dark matter}.  The spacetime
in the galactic halo region is characterized by
the line element

\begin{equation}\label{E:line2}
ds^{2}=-\left(\frac{r}{b_0}\right)^ldt^{2}
+e^{2\Lambda(r)}dr^2
+r^{2}(d\theta^{2}+\text{sin}^{2}\theta\,
d\phi^{2}),
\end{equation}
to be discussed in Sec. \ref{S:rotation}.

The purpose of this paper is to show that both
models can be obtained from the same exact
solution of the Einstein field equations given
the barotropic equation of state $p=\omega\rho$.
Moreover, for all practical purposes this
solution is the most general possible exact
solution obtainable.  The numerical value of
the parameter $\omega$ then becomes the primary
distinguisher between the two models.


\section{An exact solution}
Our first step is to list the Einstein field
equations \cite{MT88}:
\begin{equation}\label{E:Einstein1}
  \rho(r)=\frac{b'}{8\pi r^2},
\end{equation}
\begin{equation}\label{E:Einstein2}
   p_r(r)=\frac{1}{8\pi}\left[-\frac{b}{r^3}+
   2\left(1-\frac{b}{r}\right)\frac{\Phi'}{r}
   \right],
\end{equation}
and
\begin{equation}\label{E:Einstein3}
   p_t(r)=\frac{1}{8\pi}\left(1-\frac{b}{r}\right)
   \left[\Phi''-\frac{b'r-b}{2r(r-b)}\Phi'
   +(\Phi')^2+\frac{\Phi'}{r}-
   \frac{b'r-b}{2r^2(r-b)}\right],
\end{equation}
where $\rho(r)$ is the energy density, $p_r(r)$
is the radial pressure, and $p_t(r)$ the lateral
pressure.

The barotropic equation of state (EoS) $p=
\omega\rho$ has been used in various
cosmological settings, where the pressure
is necessarily isotropic.  Since this section
deals strictly with wormholes, the pressure
is now the radial pressure, leading to the EoS
\begin{equation}\label{S:EoS}
     p_r=\omega\rho,
\end{equation}
also discussed by Lobo \cite{fL06}.  The 
transverse pressure is then determined from 
Eq. (\ref{E:Einstein3}).  More importantly,
however, for the special case $\omega <-1$
discussed below, the extension to spherically
symmetric inhomogeneous spacetimes has been
carried out.  (See Ref. \cite{sS05} for
details.)

We are going to find an exact solution that
yields several special cases.
Exact solutions were discussed by Kuhfittig
\cite{pK06} and earlier by Lobo \cite{fL06}
and Zaslavskii \cite {oZ05}).  To meet the
goals in this paper, we need to find the
most general possible exact solution.  Here
it turns out to be convenient to start with
the line element
\begin{equation}\label{E:line3}
ds^{2}=-e^{2\Phi(r)}dt^{2}+e^{2\Lambda(r)}dr^2
+r^{2}(d\theta^{2}+\text{sin}^{2}\theta\,
d\phi^{2}).
\end{equation}
From line element (\ref{E:line1}) we have
$b(r)=r(1-e^{-2\Lambda(r)})$.  Substituting
Eqs. (\ref{E:Einstein1}) and (\ref{E:Einstein2})
in the EoS $p_r=\omega\rho$, we obtain
\begin{equation}\label{E:diff1}
   -\omega\Lambda'=-\Phi'+\frac{1}{2r}
   \left(e^{2\Lambda}-1\right)(\omega+1).
\end{equation}
This equation can be solved by separation of
variables only if $\Phi'\equiv 0$ ($\Phi\equiv
\text{constant})$ or if $\Phi$ is defined by
\begin{equation}\label{E:red1}
   e^{2\Phi}=\left(\frac{r}{b_0}\right)^l
   \quad\text{for any real}\,\,l,
\end{equation}
[$\Phi'=l/(2r)$], where $b_0$ is an arbitrary
constant.  The former case yields Lobo's
solution \cite{fL06}
\begin{equation}\label{E:Lobo}
   b(r)=r_0\left(\frac{r}{r_0}
   \right)^{-1/\omega}.
\end{equation}
In the latter case we have
\begin{equation*}
   -\omega\Lambda'=-\frac{l}{2r}+\frac{1}{2r}
   \left(e^{2\Lambda}-1\right)(\omega+1),
\end{equation*}
showing that Eq. (\ref{E:red1}) is the only
other choice that allows $1/r$ to be factored
out, thereby separating variables:
\begin{equation}\label{E:diff2}
   \frac{-2\omega\Lambda'}{-l+
   (e^{2\Lambda}-1)(\omega +1)}=\frac{1}{r}.
\end{equation}
The solution is
\begin{equation}\label{E:solution1}
   e^{-2\Lambda}=\frac{\omega +1}{\omega +1+l}
   +Cr^{(-\omega -1-l)/\omega}.
\end{equation}

\emph{Remark 1:} The motivation for this solution 
was clearly a mathematical one, seeking to 
obtain the most general possible exact solution.  
A natural physical interpretation is possible, 
however, for a standard perfect fluid with 
$p_r=p_t\equiv p$.  In this classical scenario, 
solutions corresponding to the linear (isothermal) 
EoS $p=\omega\rho$, $0<\omega <1$, had been 
studied earlier by Chandrasekhar \cite{sC72}.

\section{Wormhole solutions}
In this section we specialize the exact solution
(\ref{E:solution1}) to the study of wormholes.

From $b(r)=r(1-e^{-2\Lambda})$ and the requirement
$b(r_0)=r_0$, we obtain
\[
   C=\frac{\omega +1}{\omega +1+l}
   r_0^{(\omega +1+l)/\omega}.
\]
Hence
\begin{equation}\label{E:shape1}
   b(r)=\frac{l}{\omega +1+l}r+r^{(-1-l)/\omega}
   \frac{\omega +1}{\omega +1+l}
   r_0^{(\omega +1+l)/\omega}.
\end{equation}
The final result is the following exact solution:
\begin{equation}\label{E:generalized}
   e^{2\Lambda(r)}=\frac{1}{1-\frac{b(r)}{r}}=
   \frac{1}{\frac{\omega+1}{\omega+1+l}
   \left[1-\left(\frac{r_0}{r}
   \right)^{(\omega+1+l)/\omega}\right]}.
\end{equation}
When $l=-1$, we obtain Zaslavskii's form
\cite{oZ05}
\begin{equation}
   e^{2\Lambda(r)}=
   \frac{1}{\left(1+\frac{1}{\omega}\right)
   \left(1-\frac{r_0}{r}\right)}
\end{equation}
When $l=0$, we obtain Lobo's solution, Eq.
(\ref{E:Lobo}).  Both assume a phantom-energy
background, i.e., $\omega <-1$ in the EoS
$p_r=\omega\rho$.  The null energy condition
is therefore automatically violated, also
confirmed below.  The line element then
becomes
\begin{equation}\label{E:line4}
ds^{2}=-\left(\frac{r}{b_0}\right)^ldt^{2}
   +\frac{dr^2}{\frac{\omega+1}{\omega+1+l}
   \left[1-\left(\frac{r_0}{r}
   \right)^{(\omega+1+l)/\omega}\right]}
+r^{2}(d\theta^{2}+\text{sin}^{2}\theta\,
d\phi^{2}).
\end{equation}

An explicit check on the flare-out condition
$b'(r_0)<1$ is provided by
\begin{equation}\label{E:bprime1}
   b'(r)=\frac{l}{\omega +1+l}-\frac{1+l}{\omega}
   r^{(-\omega -1-l)/\omega}
   \frac{\omega +1}{\omega +1+l}
   r_0^{(\omega +1+l)/\omega}.
\end{equation}
Substituting $r=r_0$, we find that
\[
   b'(r_0)=-\frac{1}{\omega}<1
\]
only if $\omega <-1$.  This confirms the
violation of the null energy condition:
\[
   \rho +p_r=\rho +\omega\rho=(1+\omega)\rho<0.
\]
Since Eq. (\ref{E:diff1}) can only be solved 
by separation of variables only if 
$\Phi\equiv\text{constant}$ of if $\Phi$ is 
defined by Eq. (\ref{E:red1}), we have also 
shown that Eq. (\ref{E:line4}) is for all 
practical purposes  the most general possible 
exact wormhole solution given a barotropic 
equation of state.

\emph{Remark 2:} For the sake of completeness
it should be noted that we have obtained the
most general possible exact solution for
specific choices of $\Phi$ that do not depend
directly on $\Lambda(r)$ and $\Lambda'(r)$.
Ref. \cite {pK06} discusses the more abstract
form $\Phi'(r)=F[\Lambda(r)]\Lambda'(r)$ for
some function $F$ that yields an exact solution of
Eq. (\ref{E:diff1}).  However, according to Ref.
\cite {pK06}, the choices for $F$ that
simultaneously avoid an event horizon are
extremely limited and so have little bearing
on the present study.

\section{The parameter $l$}
Even though the flare-out condition is satisfied
at the throat, we still have to examine the
vicinity of the throat, in particular the
allowed range on the parameter $l$.  We first
need to observe, however, that our wormhole
spacetime is not asymptotically flat and must
therefore be cut off at some $r=r_1$ and
joined to an external vacuum solution.

Suppose we write the shape function in the
form $b(\alpha r_0)$ for $\alpha>1$, since
$r=\alpha r_0$ is a convenient measure
of the distance from the throat.  Then the
derivative becomes
\begin{equation}\label{E:bprime2}
   b'(\alpha r_0)=\frac{l}{\omega +1+l}
   -\frac{1+l}{\omega}
   \frac{\omega +1}{\omega +1+l}
   \alpha^{(-\omega -1-l)/\omega}.
\end{equation}
This form explains why there are two special
cases: if $l=-1$, then the second term is
zero and $0<b'(r)<1$, so that $b(r)<r$.
(Observe that $b'(r)$ must be greater than
zero by Eq. (\ref{E:Einstein1}).)  If $l=0$,
then the first term is zero and
\[
    b'(r)=-\frac{1}{\omega}
      \alpha^{(-\omega -1)/\omega}>0
\]
and $b'(r)<1$ since $\alpha>1$.

For the general case, the effect of the
parameter $l$ is more complicated.  Suppose we
consider the interval $r_0\le r\le r_1$, where
$r=r_1$ is the cut-off mentioned earlier.  Then
$r_1=\alpha r_0$ for some $\alpha >1$, and by
using Eq. (\ref{E:bprime2}), we can plot
$b'(\alpha r_0)$ versus $l$ (using some
arbitrary value for $\omega$), shown in Fig. 1.
\begin{figure}[tbp]
\begin{center}
\includegraphics[width=0.8\textwidth]{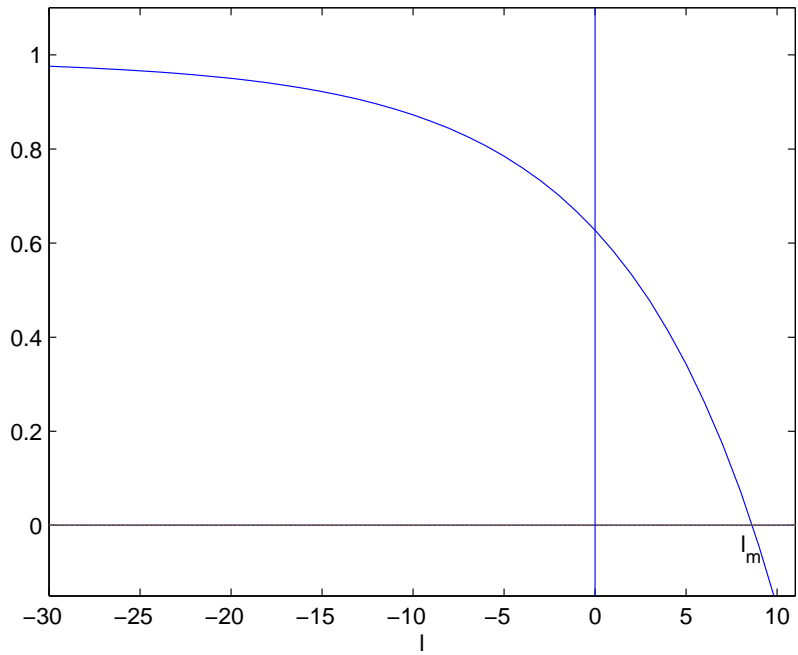}
\end{center}
\caption{$b'(r_1)$ is plotted
     versus $l$.}
\end{figure}
Its main purpose is to give a qualitative
picture of the allowed range on $l$.

According to Fig. 1, at $r=r_1$, $b'(r_1)<1$, so
that $b(r_1)<r_1$, and $b'(r_1)>0$ for all
$l\le l_m$ for some $l_m$ that depends on both
$\alpha$ and $\omega$.  For $l\le 0$, we get a
valid wormhole solution that includes both
Lobo's and Zaslavskii's solutions.  To
understand the behavior for $l>0$, consider
$b'(\alpha r_0)=0$ from Eq. (\ref{E:bprime2}).
Solving for $\alpha$, we find that $l=l_m$
is implicitly determined by
\begin{equation}\label{E:alpha}
   \alpha=\left(\frac{l_m}{l_m+1}
   \frac{\omega}{\omega +1}\right)
   ^{-\omega/(\omega +1+l_m)},
\end{equation}
which is well-defined since $\omega
/(\omega +1)>0$.  It should be noted that Fig. 1
is particularly helpful here because $l_m$
cannot be explicitly solved for.  Moreover,
the restriction $l\le l_m$ is most severe at
$r=r_1$, i.e., for any $r_2<r_1$, the above
conditions on $b'$ are automatically met on the
interval $[r_0, r_1]$, as can be seen from
Fig. 2.
\begin{figure}[tbp]
\begin{center}
\includegraphics[width=0.8\textwidth]{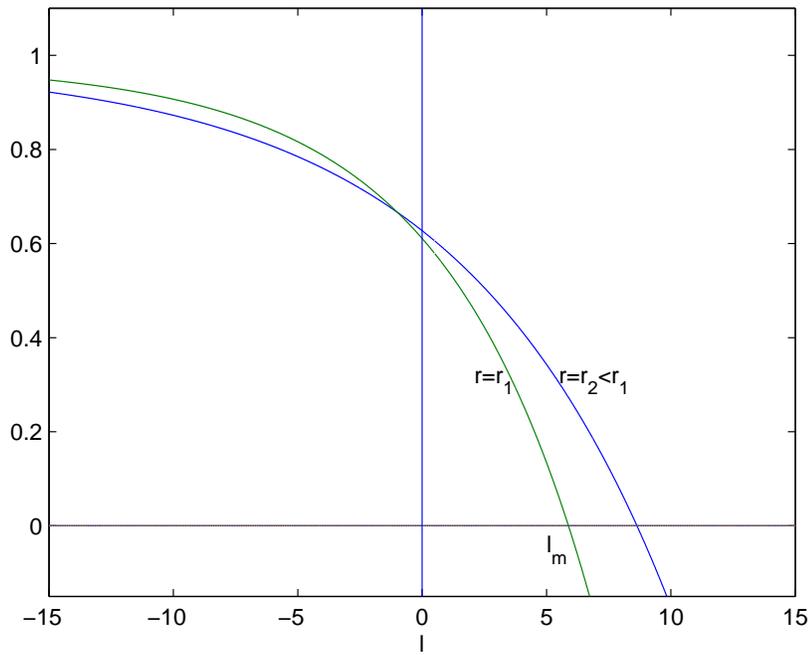}
\end{center}
\caption{$b'(r_1)$ and $b'(r_2)$, $r_1<r_2$,
     plotted versus $l$.}
\end{figure}

\section{Junction to an external vaccuum solution}
As noted in the previous section, our wormhole
spacetime is not asymptotically flat and must be
cut off at some $r=r_1$ and joined to an exterior
Scharzschild solution
\begin{equation}
ds^{2}=-\left(1-\frac{2M}{r}\right)dt^{2}
+\frac{dr^2}{1-2M/r}
+r^{2}(d\theta^{2}+\text{sin}^{2}\theta\,
d\phi^{2}).
\end{equation}
To facilitate the discussion, let us consider
the cases $l\le 0$ and $l>0$ separately,

\subsection{$l\le 0$}
The junction at the cut-off requires continuity
of the metric.  As noted in Refs. \cite{fL04, fL05},
the components $g_{\theta\theta}$ and $g_{\phi\phi}$
are already continuous due to the spherical
symmetry.  So we need to impose the continuity
requirement only on the remaining components at
$r=r_1$.  These requirements imply that
$\Phi_{\text{interior}}(r_1)=\Phi_{\text{exterior}}(r_1)$
and $b_{\text{interior}}(r_1)=b_{\text{exterior}}(r_1)$.
In particular, we must have $2M=b(r_1)$, where
$b(r)$ is given in Eq. (\ref{E:shape1}).  Also
\[
 \left(\frac{r_1}{b_0}\right)^l=1-\frac{2M}{r_1},
\]
whence
\begin{equation}\label{E:bzero}
 b_0=\frac{r_1}{(1-2M/r_1)^{1/l}}.
\end{equation}
While the metric is now continuous at the junction
surface, the derivatives may not be.  This needs to
be taken into account when discussing the surface
stresses; these are $\sigma$, the surface
stress-energy, and $\mathcal{P}$, the surface
pressure.

The following forms, proposed by Lobo \cite
{fL04, fL05}, are suitable for present purposes:
\begin{equation}\label{E:sigma}
   \sigma=-\frac{1}{4\pi r_1}\left(\sqrt
   {1-\frac{2M}{r_1}}-\sqrt{1-\frac{b(r_1)}{r_1}}
   \right)
\end{equation}
and
\begin{equation}\label{E:pressure1}
   \mathcal{P}=\frac{1}{8\pi r_1}\left(
   \frac{1-\frac{M}{r_1}}{\sqrt{1-\frac{2M}{r_1}}}
   -[1+r_1\Phi'(r_1)]\sqrt{1-\frac{b(r_1)}{r_1}}
   \right).
\end{equation}
Since $b(r_1)=2M$, the surface stress-energy
$\sigma$ is zero.  From $e^{2\Phi}=(r/b_0)^l$,
we find that $\phi'(r)=l/(2r)$ and $r_1\Phi'(r_1)
=l/2$.  So Eq. (\ref{E:pressure1}) becomes
\begin{equation*}
   \mathcal{P}=\frac{1}{8\pi r_1}\left[
   \frac{1-\frac{M}{r_1}}{\sqrt{1-\frac{2M}{r_1}}}
   -\left(1+\frac{l}{2}\right)
   \sqrt{1-\frac{b(r_1)}{r_1}}\right].
\end{equation*}
Letting $2M=b(r_1)$ and simplifying, we get
\begin{equation}\label{E:pressure2}
   \mathcal{P}=\frac{1}{8\pi r_1}
   \frac{-l+(1+l)\frac{b(r_1)}{r_1}}
   {2\sqrt{1-\frac{b(r_1)}{r_1}}}=
   \frac{1}{8\pi r_1}\frac{l\left(
   \frac{b(r_1)}{r_1}-1\right)+
   \frac{b(r_1)}{r_1}}
   {2\sqrt{1-\frac{b(r_1)}{r_1}}}>0
\end{equation}
since $l\le 0$ and $b(r_1)/r_1<1$.  Also, from
Eq. (\ref{E:Einstein2}),
\begin{equation}\label{E:radial}
   p_r(r_1)=\frac{1}{8\pi r_1^2}\left[
   -\frac{b(r_1)}{r_1}+\left(1-\frac{b(r_1)}{r_1}
   \right)l\right],
\end{equation}
it follows that $p_r(r_1)$ is negative.  Such a
combination is to be expected since a negative
radial pressure is needed to balance a positive
surface pressure.

\subsection{$l>0$}
For positive $l$, our exact solution has a
particularly interesting property: we can choose
$r=r_1$ in such a way that the surface stresses
are zero.  Such a surface is called a \emph
{boundary surface}.  To see how, we can use Eq.
(\ref{E:shape1}) to find
\begin{equation}\label{E:boundary1}
   \frac{b(r_1)}{r_1}=\frac{b(\alpha r_0)}
   {\alpha r_0}=\frac{l}{\omega +1+l}
   +\frac{\omega +1}{\omega +1+l}
   \alpha^{-(\omega +1+l)/\omega}.
   \end{equation}
If $\alpha$ is chosen so that $b'(\alpha r_0)=0$
and $l=l_m$, then we get from Eq. (\ref{E:alpha}),
\begin{equation}\label{E:boundary2}
   \frac{b(r_1)}{r_1}=\frac{l_m}{\omega +1+l_m}+
   \frac{\omega +1}{\omega +1+l_m}\left(
   \frac{l_m}{l_m+1}\frac{\omega}{\omega +1}
   \right)=\frac{l_m}{l_m+1}.
\end{equation}
It now follows from Eq. (\ref{E:pressure2}) that
$\mathcal{P}=0$.  Since we already have
$\sigma =0$, we conclude that for the case $l>0$,
we can choose the cut-off $r=r_1$ in such a way
that the junction surface is a boundary surface.
We also note that from Eq. (\ref{E:radial}),
$p_r(r_1)=0$, as expected.

\section{Galactic rotation curves}\label{S:rotation}
Returning to line element (\ref{E:line3}), if we
use the form of $e^{2\Phi}$ in Eq. (\ref{E:red1}),
we can recover line element (\ref{E:line2}),
restated here for convenience:
\begin{equation}\label{E:line5}
ds^{2}=-\left(\frac{r}{b_0}\right)^ldt^{2}
+e^{2\Lambda(r)}dr^2
+r^{2}(d\theta^{2}+\text{sin}^{2}\theta\,
d\phi^{2}).
\end{equation}
To interpret this result, let us assume that
$\omega >0$ in the equation $p=\omega\rho$.  So
we are no longer dealing with wormholes.
Instead, the EoS has a cosmological
interpretation as a perfect fluid but may
also represent dark matter.  This line element
makes physical sense only if $l>0$ (even
though our solution is mathematically correct
for any $l$), since it is normally viewed as
a model for galactic rotation curves.  Here
$l=2(v^{\phi})^2$, where $v^{\phi}$ is the
tangential velocity and $b_0$ is an (arbitrary)
integration constant \cite{NVM09, fR11}.
According to Ref. \cite{kN09}, $l=0.000001$
and remains approximately constant.

We conclude that line element (\ref{E:line5})
can be arrived at by purely mathematical means,
i.e., given the Einstein field equations and
the EoS $p=\omega\rho$, $\omega >0$, we get
an exact solution only if $l$ in Eq.
(\ref{E:red1}) is a constant.  The implication
is that a constant tangential velocity could
have been hypothesized based on the Einstein
field equations provided, of course, that a
perfect-fluid background is assumed.  The
existence of a perfect fluid is a reasonable
assumption that is also consistent with the
existence of dark matter.

\section{Conclusion}

In this paper we obtained the most general possible
exact solution of the Einstein field equations
given a barotropic equation of state.  This solution
yields two different models.  The condition
$\omega <-1$ in the anisotropic
EoS $p_r=\omega\rho$ yields the most general
possible model for wormholes supported by phantom
energy, thereby generalizing several earlier
results.  The case $\omega >0$ in the
EoS $p=\omega\rho$ yields the usual model for
galactic rotation curves. Here the EoS represents
a perfect fluid which may include dark matter.
Mathematicall speaking, we therefore have only
one exact solution, but due to the parameter
$\omega$, this solution corresponds to completely
different physical models.
\\
\\
\emph{Acknowledgment:} The author would like to
thank Vance Gladney for many helpful discussions.

\end{document}